\documentclass[reprint,aps,prl,a4paper,10pt,twocolumn,showpacs,floatfix,superscriptaddress,amsmath,amsfonts,amssymb,preprintnumbers,citeautoscript,longbibliography]{revtex4-1}
\usepackage[T1]{fontenc}
\usepackage{charter}
\usepackage[charter,greekuppercase=italicized]{mathdesign}
\usepackage[utf8]{inputenc} 
\usepackage{graphicx,color,booktabs,microtype}

\usepackage{xfrac} 
\usepackage[colorlinks,plainpages=false,linkcolor=blue,urlcolor=blue,citecolor=blue,pdfpagemode=UseNone,pdfstartview=FitBH]{hyperref}

\usepackage{mathptmx}
\usepackage[scaled=.90]{helvet}
\usepackage{courier}

\begin{document}


\title{A new magnetic phase in the nickelate perovskite TlNiO$_3$}

\author{L. Korosec}
\email[]{lkorosec@phys.ethz.ch}
\author{M. Pikulski}
\affiliation{Laboratory for Solid State Physics, ETH Z\"urich, CH-8093 Z\"urich, Switzerland}

\author{T. Shiroka}
\affiliation{Laboratory for Solid State Physics, ETH Z\"urich, CH-8093 Z\"urich, Switzerland}
\affiliation{Paul Scherrer Institut, CH-5232 Villigen PSI, Switzerland.}

\author{M. Medarde}
\affiliation{Laboratory for Scientific Developments and Novel Materials, Paul Scherrer Institut, CH-5232 Villigen, PSI Switzerland.}

\author{H. Luetkens}
\affiliation{Laboratory for Muon Spin Spectroscopy, Paul Scherrer Institut, CH-5232 Villigen PSI, Switzerland.}

\author{J. A. Alonso}
\affiliation{Instituto de Ciencia de Materiales de Madrid, CSIC, Cantoblanco, E-28094 Madrid, Spain}

\author{H. R. Ott}
\author{J. Mesot}
\affiliation{Laboratory for Solid State Physics, ETH Z\"urich, CH-8093 Z\"urich, Switzerland}
\affiliation{Paul Scherrer Institut, CH-5232 Villigen PSI, Switzerland.}

\date{\today}

\begin{abstract}

The RNiO$_3$ perovskites are known to order antiferromagnetically below a material-dependent N\'eel temperature $T_\text{N}$.
We report experimental evidence indicating the existence of a second magnetically-ordered phase in TlNiO$_3$ above $T_\text{N} = 104$\,K, obtained using nuclear magnetic resonance and muon spin rotation spectroscopy.
The new phase, which persists up to a temperature $T_\text{N}^* = 202$\,K, is suppressed by the application of an external magnetic field of approximately 1\,T.
It is not yet known if such a phase also exists in other perovskite nickelates.

\end{abstract}


\maketitle

\begin{figure}
\includegraphics[width=0.45\textwidth]{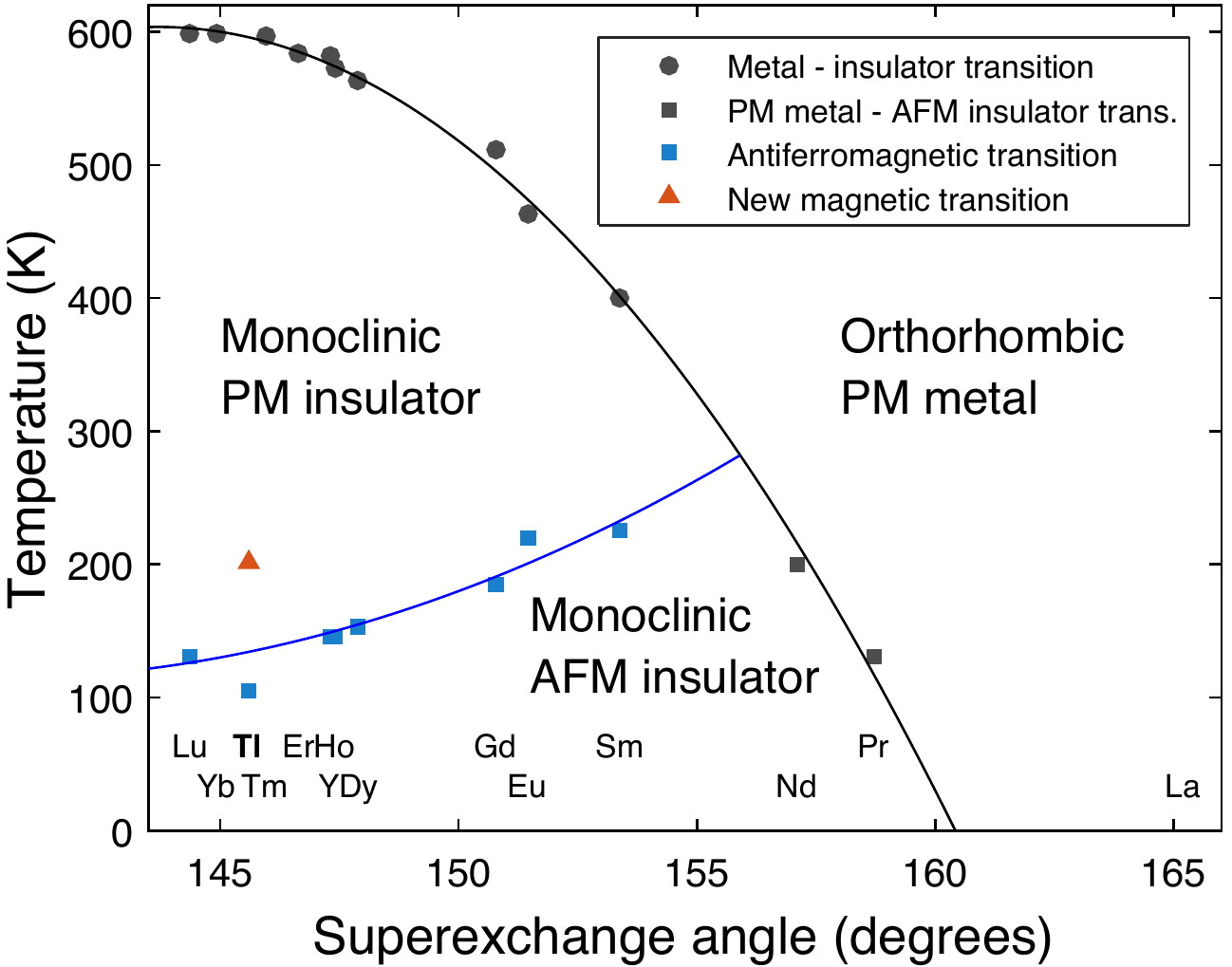}
\caption{Phase diagram of the RNiO$_3$ perovskites with respect to temperature and Ni--O--Ni superexchange angle. Data for compounds other than TlNiO$_3$ shown in this plot are taken from \cite{Medarde1997, Alonso1995, Alonso1999b, Alonso2001, Alonso2013, Alonso1999c, Munoz2009}. The red triangle indicates the new magnetic phase transition described in this work.}
\label{fig:phase_diagram}
\end{figure}

Although first synthesized already in 1970 \cite{Demazeau1971}, the rare-earth nickelates RNiO$_3$ have been the subject of intense research efforts during the last decade due to their peculiar metal--insulator transition at $T_\text{MI}$ and subsequent antiferromagnetic (AFM) order between Ni spins below a N\'eel temperature $T_\text{N}$ \cite{Medarde1997, Catalan2008}. By substituting rare-earth ions R$^{3+}$ with different radii, the perovskite lattice can be distorted continuously. This distortion affects the magnetic couplings of Ni spins by modifying the Ni--O--Ni superexchange angle.
A phase diagram of RNiO$_3$, mostly based on previously published results, is shown in Fig.~\ref{fig:phase_diagram}.
Despite recent theoretical \cite{Mazin2007, Park2012, Subedi2015} and experimental \cite{Ruppen2015} progress, the nature of the paramagnetic (PM) insulating phase is still unclear.
The AFM order below $T_\text{N}$ is characterized by a propagation vector $k_{\text{pc}} = (\sfrac{1}{4}, \sfrac{1}{4}, \sfrac{1}{4})$ with respect to the (pseudo-)cubic unit cell of the ideal perovskite structure.
Thus, a period of the magnetic structure comprises four Ni sites along each pseudocubic crystal axis.
Previous experimental reports suggested a collinear ``up-up-down-down'' structure \cite{Garcia1992, Garcia1994, Alonso1999}, while others argued for a non-collinear spiral spin configuration \cite{Fernandez2001, Scagnoli2006, Scagnoli2008, Bodenthin2011}.
The AFM phase is predicted to be ferroelectric \cite{Cheong2007, Giovannetti2009}, but this has not yet been confirmed experimentally.
Since the magnitude of the ferroelectric polarization is very different for the two candidate spin arrangements \cite{Giovannetti2009}, investigations of the magnetism of RNiO$_3$ are crucial for the understanding of their possible multiferroicity.
Nuclear magnetic resonance (NMR) is a powerful technique to locally probe magnetic behavior.
However, to our knowledge, the only previous NMR  work involving a nickelate perovskite was a $^{139}$La-NMR study of LaNiO$_3$ \cite{Sakai2002}, which is a PM metal at all temperatures.

In the present work, we present the first NMR investigation of an insulating member of the RNiO$_3$ family, in the form of a combined NMR and muon-spin rotation ($\mathrm{\mu}$SR) study of the magnetic properties of TlNiO$_3$.
This material is known to have the same qualitative behavior as the analogous rare-earth nickelates \cite{Kim2002CM}.
From an NMR perspective, TlNiO$_3$ is unique among the RNiO$_3$ compounds because both $^{203}$Tl and $^{205}$Tl nuclei are excellent for NMR.
In addition to the previously known AFM phase below $T_\text{N} = 104$\,K, our measurements reveal a previously unknown magnetically-ordered phase between $T_\text{N}$ and $T_\text{N}^* = 202$\,K.
An applied magnetic fields of 1\,T is sufficient to suppress the static magnetic order between $T_\text{N}$ and $T_\text{N}^*$.
Nevertheless, hysteretic dynamics in this temperature range are still detected by NMR, even in applied fields of several teslas (see below).

\begin{figure}
\includegraphics[width=0.45\textwidth]{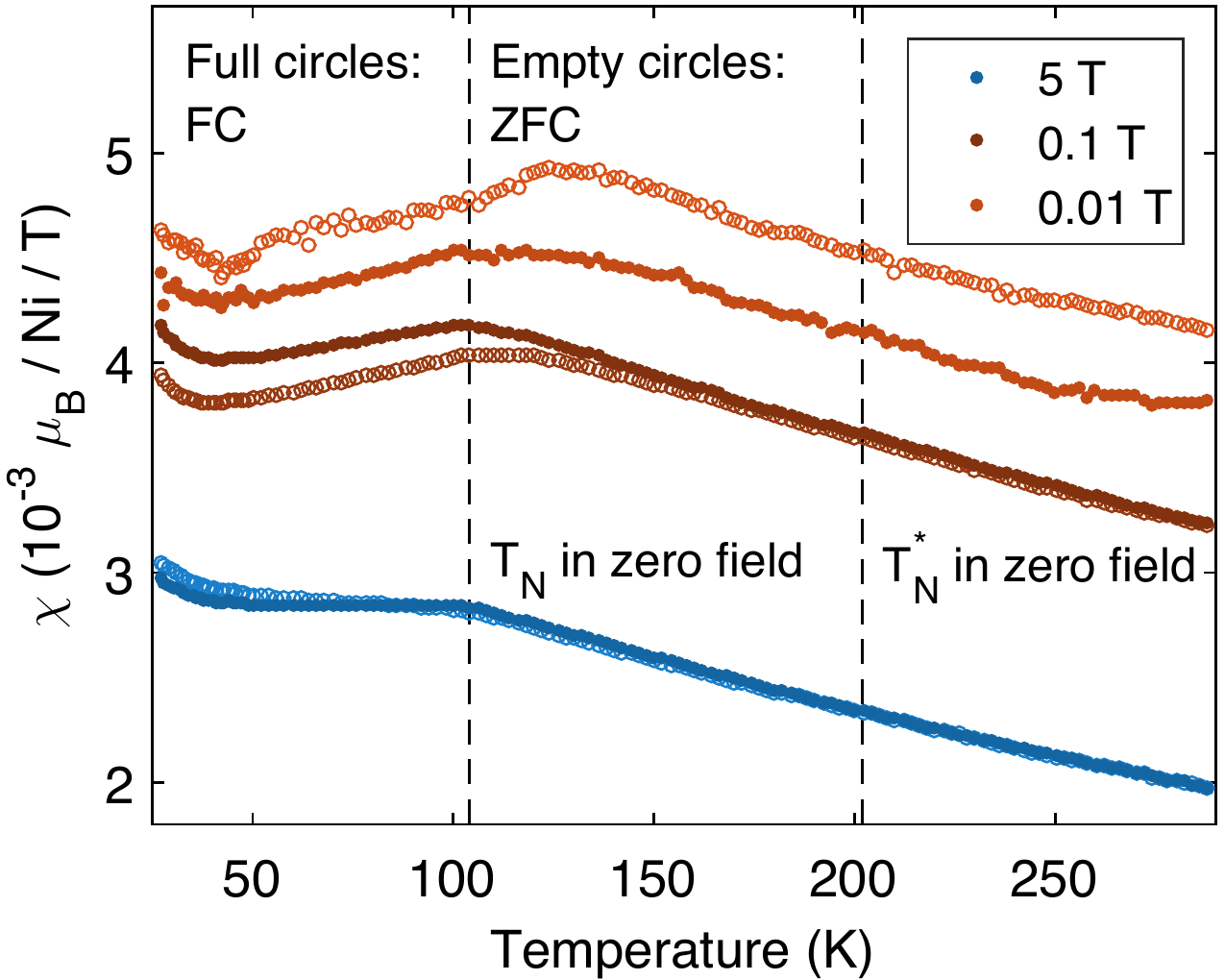}
\caption{Magnetic susceptibility of TlNiO$_3$ powder, measured using a commercial DC magnetometer at different magnetic fields. Vertical dashed lines at $T_\text{N} = 104$\,K and $T_\text{N}^* = 202$\,K indicate the magnetic transition temperatures in zero field, as determined by $\mathrm{\mu}$SR. The magnetic response below 25\,K (not shown) is dominated by a ferromagnetic impurity contribution from unreacted Ni(OH)$_2$ (see text for details).}
\label{fig:chi_vs_T}
\end{figure}

\emph{Experimental details. }The synthesis of polycrystalline TlNiO$_3$ is described in Ref.~\cite{Kim2001JMC}.
We confirmed the 99\% purity of our powder sample by X-ray diffraction.
DC magnetometry data taken at various applied fields display a kink at $T_\text{N} = 104$\,K, consistent with the previously known AFM transition (see Fig.~\ref{fig:chi_vs_T}).

At low fields, a small ferromagnetic contribution with a Curie temperature $T_\text{C} =25$\,K is observed, most likely due to the presence of residual Ni(OH)$_2$ from the synthesis \cite{Kim2001JMC}. (Note that a similar feature can be seen in the susceptibility data shown in Refs.\ \cite{Kim2001JMC, Kim2002CM} as well.)
Because of the large magnetic moment of this ferromagnet, an impurity concentration of 1\% by mass is sufficient to explain the substantial magnetic response in our data.
Due to the small mass fraction, this impurity is irrelevant for $\mathrm{\mu}$SR.
Moreover, since this impurity does not contain thallium, it does not affect our $^{203}$Tl- and $^{205}$Tl-NMR measurements, either.

\emph{Nuclear magnetic resonance. }$^{203}$Tl- and $^{205}$Tl-NMR spectra were acquired using a standard spin-echo pulse sequence.
Because of the very large linewidth, the NMR signal in the AFM phase was acquired by sweeping the frequency and integrating the Fourier-transformed spin-echo.
The transverse relaxation time $T_2$ was measured by varying the delay $\tau$ between radio-frequency pulses in the spin-echo sequence and fitting the resulting NMR amplitudes to an exponential decay $M(\tau) = M_0 \exp(-\sfrac{2 \tau}{T_2})$. Due to the rapid relaxation and decoherence rates in the PM phase ($T_1\approx 20$\,$\mu$s, $T_2\approx 16$\,$\mu$s), the spin-echo intensities are low and substantial signal-averaging is required to improve the signal-to-noise ratio of the NMR measurements.
In addition to standard NMR experiments in an applied field, we performed complementary zero-field NMR investigations in the magnetically-ordered phase.

Two stable thallium isotopes, $^{203}$Tl and $^{205}$Tl, occur naturally with abundances of 29.5\% and 70.5\%. Both nuclei have spin $\sfrac{1}{2}$ and gyromagnetic ratios of $\sfrac{\gamma_{203}}{2 \pi} = 24.3216$\,MHz/T and $\sfrac{\gamma_{205}}{2 \pi} = 24.5603$\,MHz/T, respectively. Due to the high abundances and gyromagnetic ratios, both nuclei are well suited for magnetic resonance measurements. In TlNiO$_3$, the resonance signals of both $^{203}$Tl and $^{205}$Tl were detected at a frequency shift of $K = 1.03$\% at room temperature.

\begin{figure}
\includegraphics[width=0.45\textwidth]{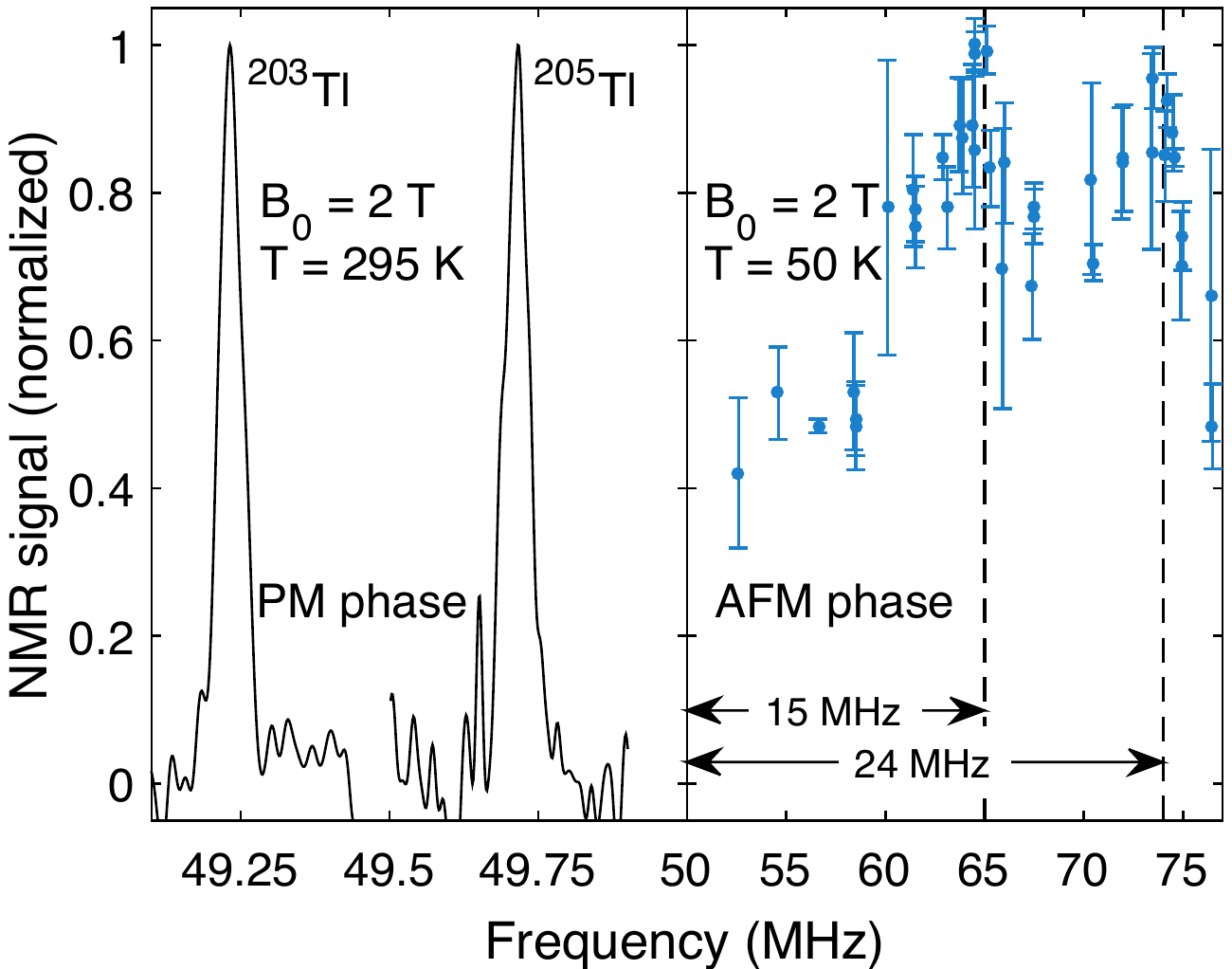}
\caption{$^{203}$Tl and $^{205}$Tl NMR spectra at 295\,K in the PM phase (left) and at 50\,K in the AFM phase (right). Note the different frequency scales in the two panels. The broadened upper edges of two overlapping AFM powder patterns are emphasized by vertical dashed lines.}
\label{fig:NMR_spectra}
\end{figure}

Due to the distribution of orientations of the intrinsic magnetic field, the NMR powder pattern broadens by two orders of magnitude in the AFM phase. Two cusps, marked by vertical dashed lines in Fig.~\ref{fig:NMR_spectra}, are seen in the NMR spectrum, which correspond to the broadened upper edges of antiferromagnetic NMR powder patterns from two magnetically-inequivalent Tl-sites.
The edges originate from crystallites whose internal magnetic field is aligned parallel to the applied field \cite{Yamada1986}. They occur at frequencies
\begin{equation}
\phantom{,} \nu_{\text{edge}} = \frac{\gamma}{2 \pi} ( B_0 + B_{\text{int}} ) \quad ,
\label{eq:AFNMR}
\end{equation}
where $B_0$ is the applied magnetic field, and $B_{\text{int}}$ is the local internal field.
This internal field is static on the typical timescale of an NMR experiment, which is at least $10^{-4}$\,s. 
Since the relative difference of the gyromagnetic ratios $\sfrac{\gamma_{205} - \gamma_{203}}{\gamma_{205}} \approx 1$\% is much smaller than the relative width of the powder pattern, the contributions from $^{203}$Tl and $^{205}$Tl cannot be distinguished in the magnetically-ordered phase. Hence, when mentioning Tl-NMR spectra in the AFM phase, we refer to the superposition of $^{203}$Tl- and $^{205}$Tl-NMR spectra.

By applying Eq.~\ref{eq:AFNMR} to the NMR spectrum acquired at 50\,K and $B_0 = 2$\,T, shown in Fig.~\ref{fig:NMR_spectra}, we extract the two internal magnetic fields $B_{\text{int}} \approx 0.6$\,T and 1.0\,T.
The zero-field (ZF) NMR frequency $\nu_\text{{ZF}}=\sfrac{\gamma}{2 \pi} B_{\text{int}}$ corresponds to $B_0 = 0$ in Eq.~\ref{eq:AFNMR}. Since $\nu_{\text{ZF}}$ is proportional to the local magnetic field, its temperature-dependence reflects the order parameter of a magnetically-ordered phase.

\begin{figure}
\includegraphics[width=0.45\textwidth]{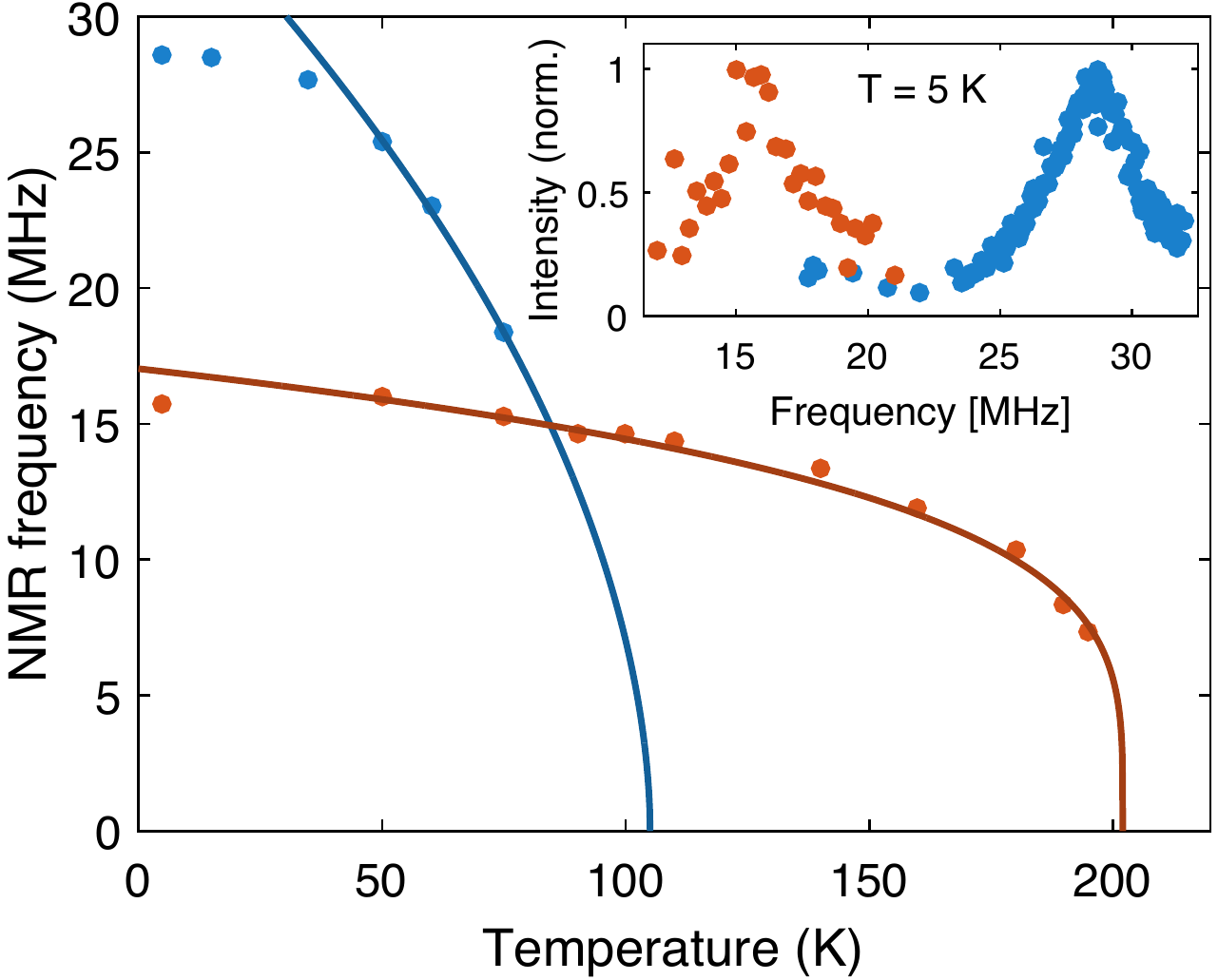}
\caption{Zero-field Tl-NMR frequencies plotted against temperature, demonstrating the presence of static internal magnetic fields up to $\sim 200$\,K. The solid lines are guides to the eye. Inset: Zero-field NMR spectra acquired at 5\,K. Two different sets of probe-head configurations and NMR pulse-sequence parameter sets were used to acquire the two parts of the spectrum, shown in blue and red. The intensities measured in two different setups are not directly comparable, hence signals are normalized to their maximal amplitude.}
\label{fig:AFNMR_freqs}
\end{figure}

By following the temperature-dependence of the two ZF-NMR frequencies, as plotted in Fig.~\ref{fig:AFNMR_freqs}, we identify two magnetic phases --- the previously reported phase below $T_\text{N} = 104$\,K \cite{Kim2001JMC, Kim2002CM} and a new phase which persists up to $T_\text{N}^*\sim200$\,K. We were unable to establish the exact transition temperature from ZF-NMR alone, because the NMR signal-to-noise ratio deteriorates significantly at low resonance frequencies, close to the transition. 

\begin{figure}
\includegraphics[width=0.45\textwidth]{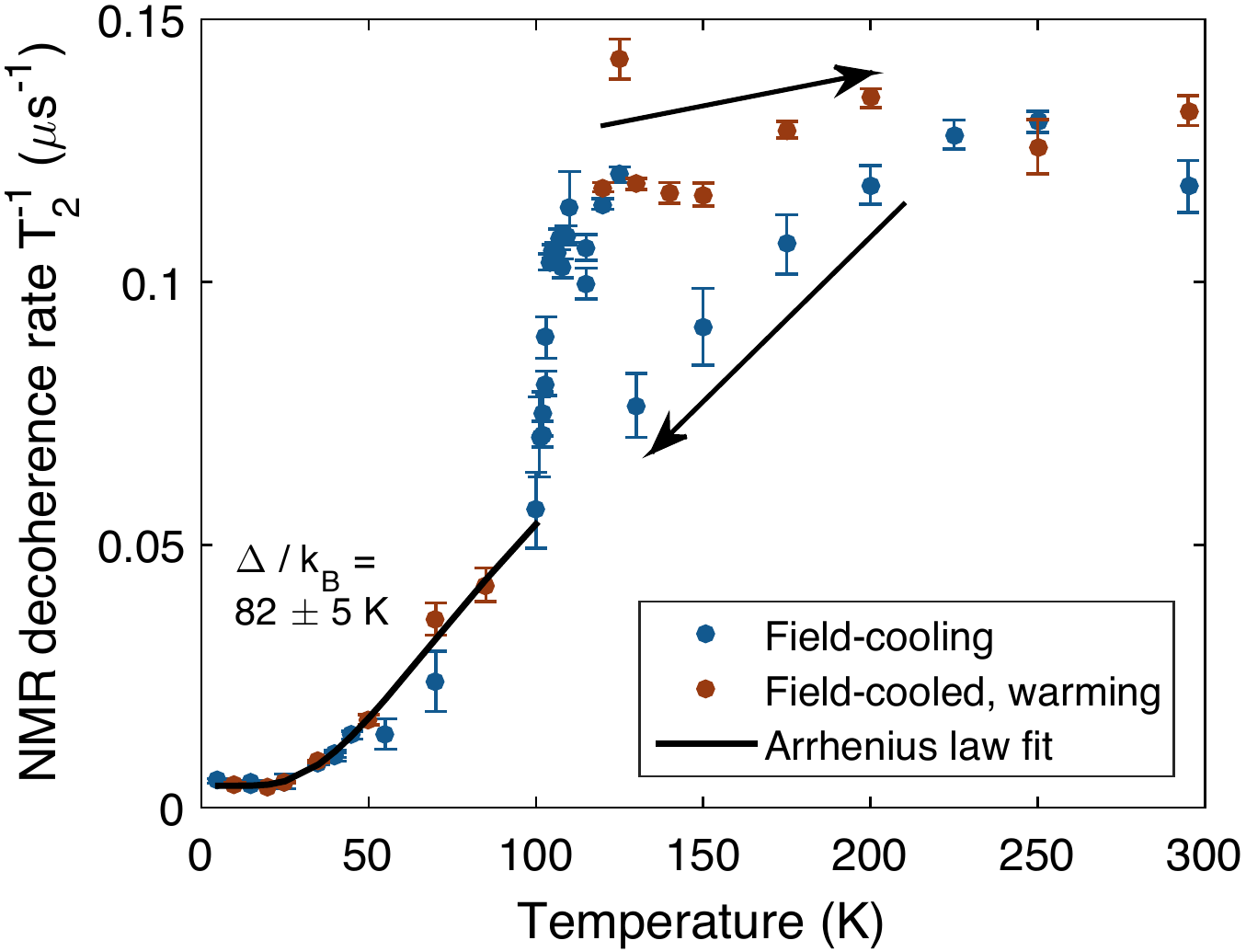}
\caption{\label{fig:NMR_T2}Transverse NMR relaxation rate $T_2^{-1}$ in an applied magnetic field of several teslas, plotted against temperature. Note the large hysteresis between 125--200\,K. Below 100\,K, the relaxation rate follows an Arrhenius law with an activation energy 
${\sfrac{\Delta}{k_\text{B}}=82 \pm 5}$\,K.}
\end{figure}

Measurements performed in $B_0=3.5$\,T did not reveal any static magnetic order above $T_\text{N}$.
However, as shown in Fig.~\ref{fig:NMR_T2}, the dephasing rate of the nuclear spins $T_2^{-1}$ shows a significant hysteresis between $\sim 125$\,K and 200\,K. This almost coincides with temperature range where we find the new magnetic phase in zero field, between $T_\text{N}$ and $T_\text{N}^*$.
Below 100\,K, both the dephasing rate $T_2^{-1}$ and the relaxation rate $T_1^{-1}$ (not shown) follow an Arrhenius law $T_i^{-1} = \tau_i \, \exp(-\sfrac{\Delta}{k_\text{B} T}) + C_i$ with an energy gap $\sfrac{\Delta}{k_\text{B}} = 82 \pm 5$\,K for the magnetic excitations.

\begin{figure}
\includegraphics[width=0.45\textwidth]{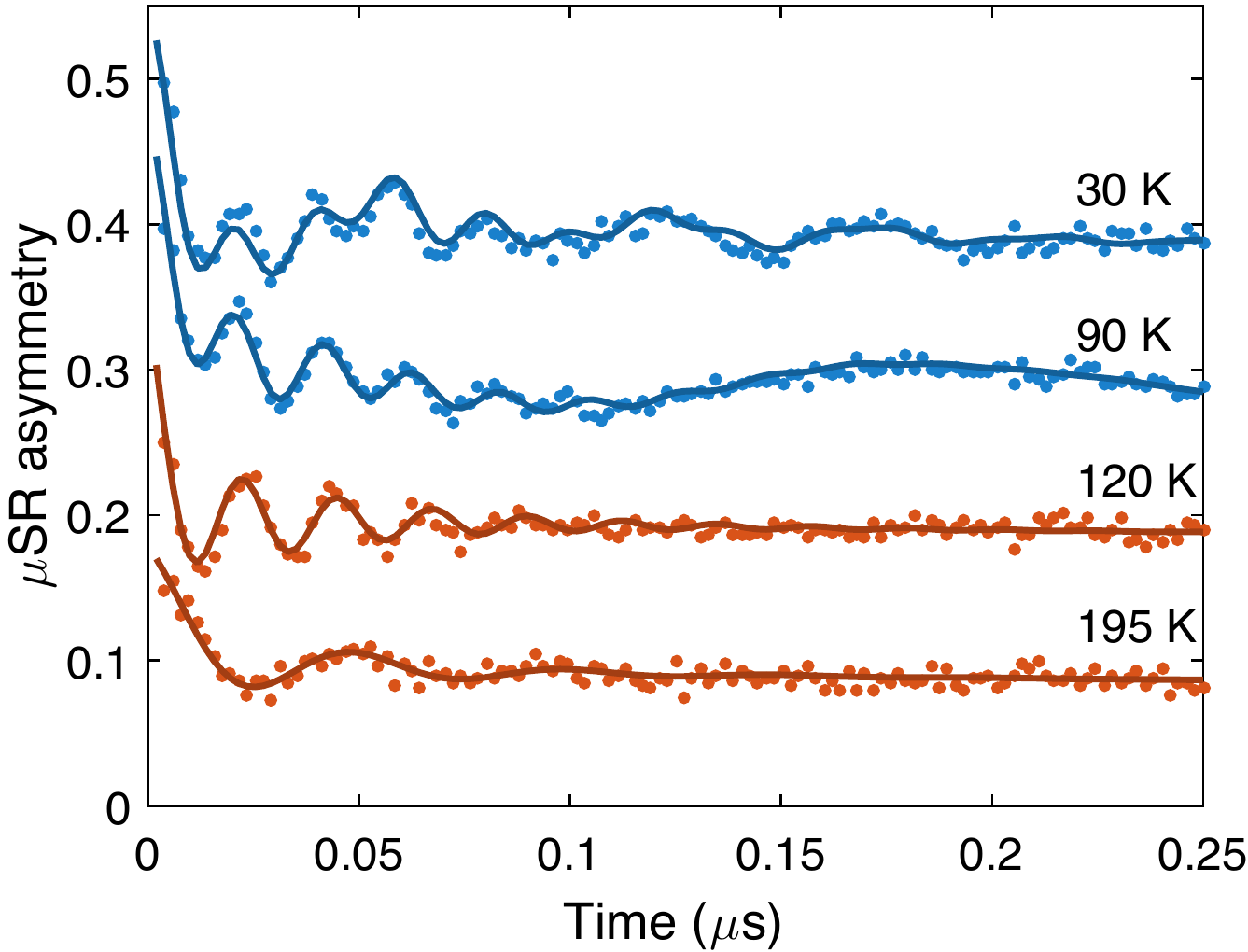}
\caption{Comparison of experimental zero-field $\mu$SR asymmetries to the fits (solid line) described in the text body. Note that the oscillations mostly dephase within 0.2\,$\mu$s, corresponding to a strong broadening of the frequency spectrum. Data acquired at different temperatures are offset vertically for clarity.}
\label{fig:muSR_histograms}
\end{figure}

\emph{Muon-spin rotation. }Muon-spin rotation ($\mu$SR) experiments were performed in TlNiO$_3$ powder in zero applied field between 5\,K and ambient temperature using the General Purpose Spectrometer (GPS) at the Swiss Muon Source (S$\mu$S) at the Paul Scherrer Institut.  
In analogy to ZF-NMR, the zero-field $\mathrm{\mu}$SR frequencies reflect the local internal magnetic field (at the muon stopping site), hence providing a measure of the order parameter in the magnetically-ordered phase. Unlike NMR, $\mu$SR can detect low-frequency signals without loss of amplitude, as can be seen in Fig.~\ref{fig:muSR_histograms}.
This allows for the determination of the new phase transition temperature $T_\text{N}^* = 202$\,K in zero field.

The time-domain histogram of detector $i$, $S_i (t)$ is given by 
\begin{equation}
\phantom{,}S_i (t) = N_{0i} \, \big[1 + \operatorname{Re} \, A(t) \, \mathrm{e}^{\mathrm{i}\phi_i} \big] \, \mathrm{e}^{-\frac{t}{\tau_\mu}} + B_i \quad,
\label{eq:histogram}
\end{equation}
where $A(t)$ is the asymmetry, $N_{0i}$ are normalized signal count rates, $\phi_i$ are phase offsets, $B_i$ are background count rates, and $\tau_\mu = 2.2$\,$\mu$s is the muon lifetime \cite{Yaouanc2011}.

First, the $\phi_i$ are obtained from a calibration measurement in the PM phase in a transversely applied field of 3\,mT. Second, the $N_{0i}$ and $B_i$ are fitted for each dataset using Eq.~\ref{eq:histogram} and assuming $A(t)$ to be constant, which is a good approximation at long times $t \gtrsim 0.3$\,$\mu$s.
Then, the parameters of $A(t)$ are estimated by means of a global maximum-likelihood fit of the detector histograms.
The Python \cite{Python_v2p7p11} package iminuit \cite{iminuit_v1p2} is used to perform the fits and determine the standard errors using the Minos algorithm \cite{minuit}.

\begin{figure}
\includegraphics[width=0.45\textwidth]{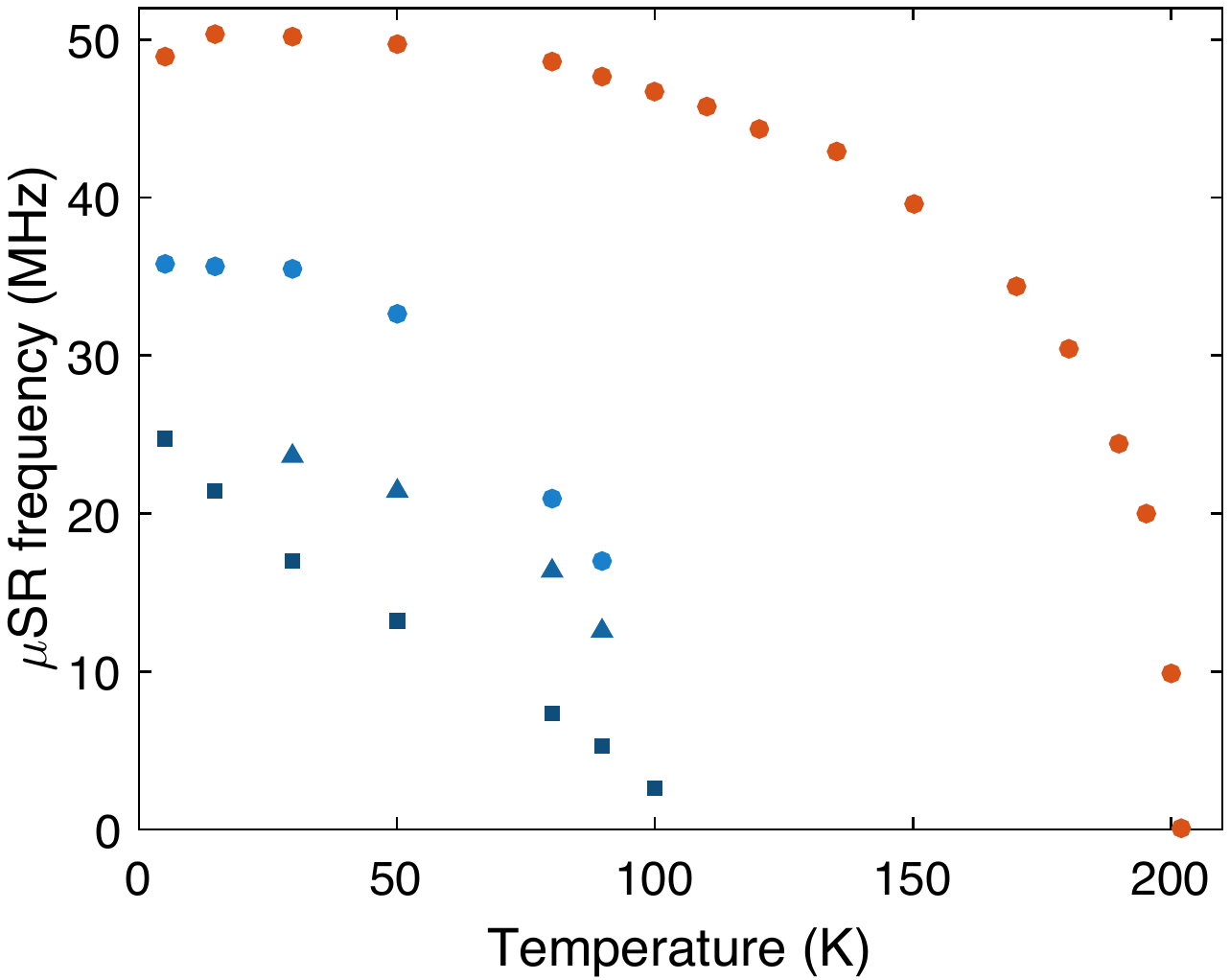}
\caption{Zero-field $\mu$SR frequencies $\nu^{(i)}$ in TlNiO$_3$ powder, as obtained from Eqs.~\ref{eq:asymmetry_highT} and \ref{eq:asymmetry_lowT}. Parameters were fitted with a maximum-likelihood method using Minuit \cite{minuit}. Different markers are assigned to different components of the $\mathrm{\mu}$SR asymmetry. Error bars are omitted because they are smaller than the symbol size.}
\label{fig:muSR_freqs}
\end{figure}

Above $T_\text{N}$, the expression
\begin{equation}
A(t) = A \, \Big[\tfrac{2}{3} \cos (2 \pi \nu^{(1)} t) \mathrm{e}^{-\lambda_\text{x}^{(1)} t} + \tfrac{1}{3} \mathrm{e}^{-\lambda_\text{z} t}\Big] + B \, \mathrm{e}^{-\lambda_{\text{DC}} t} + C
\label{eq:asymmetry_highT}
\end{equation}
is used to fit the $\mu$SR asymmetry \cite{Yaouanc2011, Reotier1997}.
This expression accounts for two muon stopping sites: one where the mean local magnetic field is non-zero and one where it is zero; at both sites exponential relaxation occurs by fluctuations of the magnetic field.
Exponential relaxation was also found in previous $\mu$SR experiments on other RNiO$_3$ perovskites \cite{Garcia1995, Garcia2006}.

Below $T_\text{N}$, the asymmetry is modeled using
\begin{equation}
\phantom{,}A(t) = a^{(0)} \mathrm{e}^{-\lambda_\text{z} t} + \sum _{i=1}^4 a^{(i)} \cos (2 \pi \nu^{(i)} t) \mathrm{e}^{- \lambda_\text{x}^{(i)} t} + C\quad,
\label{eq:asymmetry_lowT}
\end{equation}
which describes the $\mu$SR signal from four magnetically-inequivalent muon sites experiencing non-zero local fields and exponential relaxation \cite{Yaouanc2011, Reotier1997}.
A comparison between the experimental asymmetry and our fits is shown in Fig.~\ref{fig:muSR_histograms}. Since the $\mu$SR signal dephases within the first 0.2\,$\mu$s, the Fourier-transformed spectrum is strongly broadened. Thus it is impossible to resolve oscillations whose frequencies differ by less than a few MHz. This is the reason why we could establish only two frequencies at 100\,K, and only three frequencies at 5\,K and 15\,K. In the corresponding fits, a rapidly relaxing DC component was added phenomenologically to improve the fit convergence.

The temperature-dependence of the frequencies $\nu^{(i)}$ plotted in Fig.~\ref{fig:muSR_freqs} clearly shows the two phase transitions at $T_\text{N} = 104$\,K and $T_\text{N}^* = 202$\,K. Additionally, one can see that there are four magnetically inequivalent muon sites below $T_\text{N}$, giving rise to four different oscillation frequencies. The physical implication of these four muon sites is an open question. Since there are three inequivalent O$^{2-}$-sites in TlNiO$_3$ \cite{Kim2002CM}, a tentative explanation of the three frequencies appearing below $T_\text{N}$ is that they originate from muons bound to the three different oxide ions. We have not been able to explain the approximately linear temperature-dependence of the lowest $\mu$SR frequency. However, this may be related to the occurence of a magnetically ordered phase embedded within another ordered phase.

\emph{Conclusion. }Our NMR and $\mu$SR data reveal the presence of a not-yet-reported magnetically-ordered phase in TlNiO$_3$ between $T_\text{N} = 104$\,K and $T_\text{N}^* = 202$\,K in low magnetic fields and confirm the previously-known AFM phase below $T_\text{N}$.
Both phases clearly show static magnetic order on the timescales of NMR and $\mu$SR.
Due to the strong broadening of both the NMR and $\mu$SR spectra, a distinction between short-range and long-range order is not possible,.
Future $\mu$SR experiments on TlNiO$_3$ are intended to map the phase boundary of the newly reported magnetic phase as a function of applied magnetic field and temperature, and to explore the significance of the four $\mu$SR frequencies identified below $T_\text{N}$. 
Further measurements on other compounds will show whether this new phase is unique to TlNiO$_3$ or a universal feature of the RNiO$_3$ family.

\begin{acknowledgments}
This work was financially supported in part by the Schweizerische Nationalfonds zur F\"{o}rderung der Wissenschaftlichen Forschung (SNF).
 JAA acknowledges the Spanish MINECO for funding the project MAT2013-41099-R.
\end{acknowledgments}


%

\end{document}